\documentclass[journal=jacsat,manuscript=article]{achemso}
\graphicspath{{./},{./Figures/}}
\usepackage{chemformula} % Formula subscripts using \ch{}
\usepackage[T1]{fontenc} % Use modern font encodings

\usepackage[separate-uncertainty = true]{siunitx}% package for SI units
\author{Julia J. Goedecke}
\email{julia.goedecke@physnet.uni-hamburg.de}
\affiliation[Hamburg University]
{Department of Physics, University of Hamburg, D-20355 Hamburg, Germany}

\author{Lucas Schneider}
\affiliation[Hamburg University]
{Department of Physics, University of Hamburg, D-20355 Hamburg, Germany}

\author{Yingqiao Ma}
\affiliation[Hamburg University]
{Department of Physics, University of Hamburg, D-20355 Hamburg, Germany}
\alsoaffiliation[CAS]
{Present address: Beijing National Laboratory for Molecular Sciences, CAS Key Laboratory of Organic Solids, Institute of Chemistry, Chinese Academy of Sciences, Beijing 100190, China}

\author{Khai Ton That}
\affiliation[Hamburg University]
{Department of Physics, University of Hamburg, D-20355 Hamburg, Germany}

\author{\\Dongfei Wang}
\affiliation[Hamburg University]
{Department of Physics, University of Hamburg, D-20355 Hamburg, Germany}
\alsoaffiliation[Nanogune]
{Present address: CIC Nanogune, Tolosa Hiribidea, 76
E-20018 Donostia – San Sebastian}

\author{Jens Wiebe}
\email{jwiebe@physnet.uni-hamburg.de}
\affiliation[Hamburg University]
{Department of Physics, University of Hamburg, D-20355 Hamburg, Germany}

\author{Roland Wiesendanger}
\affiliation[Hamburg University]
{Department of Physics, University of Hamburg, D-20355 Hamburg, Germany}

\title[XXX]
  {Correlation of Magnetism and Disordered Shiba bands in Fe Monolayer Islands on Nb(110)}

\abbreviations{IR,NMR,UV}

\begin{document}

\begin{abstract}
Two-dimensional (2D) magnet-superconductor hybrid systems are intensively studied due to their potential for the realization of 2D topological superconductors with Majorana edge modes. It is theoretically predicted that this quantum state is ubiquitous in spin-orbit coupled ferromagnetic or skyrmionic 2D spin-lattices in proximity to an $s$-wave superconductor. However, recent examples suggest that the requirements for topological superconductivity are complicated by the multi-orbital nature of the magnetic components and disorder effects. Here, we investigate Fe monolayer islands grown on a surface of the $s$-wave superconductor with the largest gap of all elemental superconductors, Nb, with respect to magnetism and superconductivity using spin-resolved scanning tunneling spectrosopy. We find three types of Fe monolayer islands which differ by their reconstruction inducing disorder, the magnetism and the sub-gap electronic states. All three types are ferromagnetic with different coercive fields indicating diverse exchange and anisotropy energies. On all three islands, there is finite spectral weight throughout the substrate's energy gap at the expense of the coherence peak intensity, indicating the formation of Shiba bands overlapping with the Fermi energy. The gap filling and coherence peak reduction is strongest for the island with largest coercive field. A strong lateral variation of the spectral weight of the Shiba bands signifies substantial disorder on the order of the substrate's pairing energy with a length scale of the period of the three different reconstructions. There are neither signs of topological gaps within these bands nor of any kind of edge modes. Our work illustrates that a reconstructed growth mode of magnetic layers on superconducting surfaces is detrimental for the formation of 2D topological superconductivity. 
\end{abstract}

\section{Introduction}
 Two-dimensional (2D) chiral \emph{p}-wave, or more generally \emph{topological}, superconductors are predicted to host exotic dispersive 1D electron states on their edges, named Majorana edge modes~\cite{Qi2010}. Such systems recently attracted a lot of attention because of their potential for the realization of concepts for fault tolerant quantum computation~\cite{Kitaev2003}. One of the most prominent proposals for the realization of such topological superconductors are  Shiba-lattices, i.e. 2D lattices of magnetic atoms, or thin magnetic layers, deposited on the surface of a conventional \emph{s}-wave superconductor which supplies strong spin-orbit coupling.~\cite{Roentynen2015, Roentynen2016, Li2016, Rachel2017} Related concepts which eliminate the need of strong spin-orbit coupling~\cite{Zlotnikov2021} propose the utilization of non-collinear spin structures\cite{Nakosai2013,Chen2015,Zlotnikov2021}, in particular Skyrmions \cite{Pershoguba2016, Yang2016, Mohanta2021, Mascot2021} or Skyrmion-like spin-structures\cite{Garnier2019} coupled to conventional \emph{s}-wave superconductors.

Experimental investigations of 2D Shiba-lattices so far focused on magnetic transition metal islands on superconducting crystals with a thin oxide decoupling layer~\cite{Palacio-Morales2019}, van-der-Waals heterostructures~\cite{Kezilebieke2020, Kezilebieke2022}, transition metal-silicon alloys buried below superconducting layers~\cite{Menard2017,Menard2019}, spin-spirals~\cite{Mougel2022} or antiferromagnetic transition metal layers on elemental superconductors~\cite{LoConte2021}. The reported experimental evidences of topological superconductivity in some of these systems~\cite{Palacio-Morales2019, Kezilebieke2020, Kezilebieke2022, Menard2017,Menard2019} relied on the detection of zero bias resonances or an enhanced density of states localized on the rim of the lattice. However, it was theoretically predicted that the Majorana edge modes have a dispersion, which can be experimentally resolved~\cite{Wang2020}. The realization of Shiba-lattices directly on a surface of a metal usually enables atomic manipulation~\cite{Schneider2022, Kuester2021b} and, thereby, would permit the building of more complex structures, as, e.g. 1D Shiba-chains attached to 2D Shiba-lattices allowing for additional experiments in order to prove or disprove the Majorana edge mode origin of enhanced densities of states~\cite{Mascot2019}.

Yet, studies of transition metal layers in \emph{direct} contact to elementary superconductors did not show indications for topological superconductivity and dispersing Majorana edge modes~\cite{Palacio-Morales2019}. For such lattices in direct contact to the superconductor, the realization of a SC state in the ferromagnet by the proximity effect~\cite{Buzdin2005,Bergeret2005,Linder2009} can be hampered by a strong inverse proximity effect, which may reduce the magnetism~\cite{Muehge1998, Muehge1998b} and might also quench the superconductivity in the vicinity of the Shiba-lattice (see citations in~\cite{Linder2009}). Another important issue for the development of 2D topological superconductivity is disorder. It has been shown theoretically, that disorder can have strong effects on the topologically superconducting properties of 1D \cite{Kim2014, Hui2015, Weststroem2016, Awoga2017} and 2D~\cite{Mascot2019b, Lu2020, Christian2021} Shiba-lattices. Considering potential disorder in the superconductor, it has been shown for the case of a ferromagnetic spin chain (1D)~\cite{Awoga2017}, that the topological minigap is very sensitive to disorder, but the MM is surprisingly insensitive, as long as the surrounding superconductor does not show signs of strong disorder. For the 2D case, e.g. induced in a layer with a Rashba-type spin-orbit coupling sandwiched between a ferromagnet and a superconductor~\cite{Lu2020}, very weak potential disorder first decreases the Majorana bound state localization length with increasing disorder strength. However, it increases again when the disorder gets on the order of the pairing energy. This sensitivity can get even stronger for correlated disorder~\cite{Christian2021} and is particularly effective if the correlation length of the disorder in the chemical potential is comparable with the coherence length of the superconductor. Magnetic disorder was investigated for the 1D spin-chain system, e.g. induced by missing magnetic atoms or variations in the Shiba couplings~\cite{Weststroem2016}, and showed that the topological phase is relatively sensitive to such kinds of disorder. For the 2D system~\cite{Mascot2019b}, magnetic disorder either in the form of a disordered magnetic coupling between a ferromagnetic lattice and the electrons in the superconductor, for randomly oriented spins, or for randomly missing spins, more strongly suppresses the topological phases as compared to potential disorder. Here, the latter type of magnetic disorder leads to the strongest suppression of topological phases, but correlated disorder has a weaker effect. On the other hand, magnetic disorder may also favor topological phases in 2D Shiba-lattices, e.g. for very particular values of the chemical potential~\cite{Mascot2019b}, for Moir\'e-lattices~\cite{Kezilebieke2022}, or for a spin-glass state~\cite{Poeyhoenen2018}.
Therefore, it is surprising, that disorder effects in 1D~\cite{Kamlapure2018, Liebhaber2021} or 2D Shiba-lattices~\cite{Menard2017,Menard2019, Kezilebieke2022} have been barely touched on experimentally.

Here, we experimentally investigate ferromagnetic 2D Shiba-lattices of the transition metal Fe grown directly ontop of the clean (110) surface of Nb, the elemental superconductor with the largest energy gap. We find that the system realizes strongly disordered Shiba-bands induced by ferromagnetic Fe monolayer reconstructions, and study these bands with respect to indications for topological superconductivity.

\section{Results and Discussion}

\subsection{Reconstructions of the Fe monolayer grown on clean Nb(110)}
The growth of Fe on the (110) surface of Nb films deposited on various substrates has previously been intensely studied by RHEED, Auger-electron spectroscopy, LEED and STM~\cite{Mougin2002, Wolf2006}. For room temperature deposition of very thin Fe films, which is also the case studied in the current work, there are no signs of intermixing between Fe and Nb~\cite{Mougin2002, Wolf2006}. Here, we grow thin films of Fe on the clean oxygen-reconstruction-free (110) surface of a Nb single crystal (Figure~\ref{fig1}a,b) which is achieved by carefully flashing the crystal in ultra-high vacuum to very high temperatures (see Methods). Note, that for previous studies, the substrate most probably was still at least partly oxygen-reconstructed resulting in a different Fe structure compared to the results we present here (see Supplementary Note~1 and Figure~S1). After RT evaporation of Fe with less than a monolayer (ML) coverage onto such clean Nb(110) surfaces, three different reconstructions were observed on ML Fe islands (see Figure~\ref{fig1}a) with a typical height of \SI{250}{pm} (Figures~\ref{fig3} and \ref{fig4}d below), referred to as type I, II and III in the following. In addition, a few double-layered (DL) Fe islands (bright stripe) can be found, which are elongated along the [001]-direction. All three Fe ML reconstructions are found within single islands but also on separate islands. Figures~\ref{fig1}b-e show atomically resolved STM images of the Nb(110) substrate and of the individual reconstructions. In the following, we will always use the same color code in order to indicate the different types of reconstruction in the Figures: blue for type I, orange for type II, and green for type III. In Figure~\ref{fig1}b, the atomic lattice of Nb(110) is visible and the surface unit cell is marked. For comparison, this unit cell is added to the STM images of the three Fe ML reconstructions in Figures~\ref{fig1}c-e. The reconstructions I and III span a large number of substrate unit cells. Type I (Figure~\ref{fig1}c) consists of 1 to 2 nm long stripes with a larger apparent height arranged on a roughly $8 \times 16$ superlattice (with respect to $[001] \times [1\bar{1}0]$) separated by rather flat areas. Type III (Figure~\ref{fig1}e) consists of units that span $4 \times 3$ to $7 \times 4$ unit cells interrupted by dislocation lines along $[001]$. The arrangemens of both of these reconstructions are strongly influenced by the island's shape. Reconstruction II spans only 4 unit cells and resembles a ($2 \times 2$) superlattice where every second row along $[001]$ is missing. We refrain, here, from the determination of the atomic structure within these three reconstructions which would require a detailed comparison of the STM images to first principles calculations.

Spectra on the Nb(110) substrate show a distinct peak at $V \approx$~\SI{-450}{mV} (dashed line in Figure~\ref{fig1}f), which has previously been identified as the signature of the $d_{z^2}$-type surface state of Nb(110)\cite{Odobesko2019}, providing additional evidence that the Nb substrate is indeed largely clean and free from oxygen. Its spectral signature is still present, yet strongly suppressed, for the type I Fe ML reconstruction, but it is barely visible as a shoulder for the type II reconstruction and completely suppressed for the type III reconstruction. Instead all Fe ML reconstructions display a peak at $V \approx$~+\SI{400}{mV} (dashed line in Figure~\ref{fig1}f) which is absent on Nb(110). After the identification of the three types of Fe ML reconstructions grown on the clean Nb(110) we continue with the investigation of their magnetic properties.

\subsection{Spin orders of the Fe ML reconstructions}
To investigate the magnetism of the three types of Fe MLs, spin-resolved scanning tunneling spectroscopy (SP-STS) has been performed. Using a spin-polarized STM tip (see Methods), spectra taken on different regions of the reconstructions for varying out-of-plane magnetic fields within a bias range of $V$~=~$\pm$\SI{1}{V} (Supplementary Note~2 and Figure~S2) reveal the particular bias voltages that have been used in the following in order to achieve maximum spin-contrast in spin-resolved d$I$/d$V$ maps. To emphasize the magnetic contrast over electronic contributions, asymmetry maps were calculated (see Methods, Eq. (\ref{Eq:1})). Examples of such spin asymmetry maps for $B_{z_{1,2}}$~=~$\pm$\SI{0.5}{T} taken from one of each type of the Fe ML islands shown in Figure~\ref{fig2}d can be seen in Figure~\ref{fig2}a-c. It is apparent that the type I ML exhibits the strongest spin contrast, followed by type II and type III. For type I, a contrast reversal between flat and stripe regions can be observed, which we will discuss below. Hysteresis loops extracted from out-of-plane magnetic field dependent d$I$/d$V$ maps (see Methods) by averaging over a certain region on top of each type of Fe ML island are shown in Figure~\ref{fig2}e. They reveal a butterfly shaped hysteresis for type I and type II with a strong change in the spin-resolved signal, while there is hardly any change at first sight on the type III Fe ML. Since, for these measurements, only a small amount of magnetic Fe material was transferred onto the Nb tip (see Methods), it is expected that the tip is soft magnetic and aligns parallel to the external magnetic field already at several hundreds of mT~\cite{Wiebe_2011}. This is indeed evidenced by the hysteresis loops measured on type I and II islands. After an initial increase in the d$I$/d$V$ signal for the first few hundreds of mT up to $B_z$~=~$\pm$\SI{0.5}{T}, a contrast difference relative to $B_z$~=~\SI{0}{T} can be seen for all island types I and II, regardless of their size. In contrast, assuming a constant tip magnetization for this magnetic field regime would imply that all islands switch their magnetization at a similar, very small coercive field, which is highly unlikely. We, therefore, conclude that for $|B_z| >$~\SI{0.5}{T}, the tip magnetization points down (up, w.l.o.g.) for negative (positive) magnetic field in Figure~\ref{fig2}e. This is also schematically shown in Figure~\ref{fig2}f. After this characterization of the tip magnetization, which is the same for the measurements on all three ML types in Figure~\ref{fig2}e, we can now interpret the rest of the hysteresis loops, starting with the type I ML (Figure~\ref{fig2}e, blue data). For this type, d$I$/d$V$ values were averaged over flat as well as stripe regions (see Supplementary Note~2 and Figure~S2), with only the results over the flat regions shown here. For the downwards sweep (along the direction of the horizontal arrow), the d$I$/d$V$ signal first increases between $B_z$~=~\SI{0}{T} to $B_z$~=~\SI{-0.5}{T} due to the downwards reorientation of the tip magnetization, and then stays constant until about -2T. Between -2T and -2.4T the d$I$/d$V$ signal suddenly decreases. When the magnetic field is swept up again, the d$I$/d$V$ signal first increases between -0.5T and +0.5T due to the upwards reorientation of the tip magnetization, then stays constant up to about +2T, and then suddenly decreases again between +2T and +2.4T. We thus assign the changes in the d$I$/d$V$ signal between $\pm$\SI{2}{T} and $\pm$\SI{2.4}{T} to the change in the magnetization of the type I island (see the sketch in Figure~\ref{fig2}f). Exactly the same behaviour of the hysteresis loop is observed on the stripe regions, just with the inverted spin contrast that was already implied by the spin asymmetry map in Figure~\ref{fig2}a (see Supplementary Note~2 and Figure~S2). We neither observe any indications for non-collinear spin-structures,~\cite{Bode2007} nor a shifting in the maximum spin contrast between the stripe and flat regions which would imply unreasonably small domains (Supplementary Note~3 and Figure~S3). Therefore, we conclude that this type I reconstruction island has a single-domain out-of-plane ferromagnetic spin-order with a coercive field of \SI{2.2}{T}$~\pm~$\SI{0.2}{T}. The sign-changes in the spin-asymmetry (Figure~\ref{fig2}a) are most probably due to an inversion of the vacuum spin-polarization above the stripes with respect to the flat regions (Supplementary Note~2 and Figure~S2)~\cite{Zhou2010}.

The magnetization curve measured on the type II island (Figure~\ref{fig2}e, orange data points) shows qualitatively the same behaviour indicating a single domain out-of-plane ferromagnetic spin order also in this case. However, here the coercive field is considerably smaller (\SI{0.75}{T}$~\pm~$\SI{0.25}{T}). For the magnetization curve measured on the type III island, the d$I$/d$V$ signal stays largely constant for the entire magnetic field range, apart from some tiny signal changes in the small field regime where the tip magnetization direction is reoriented. This also holds true if we use bias voltages in the entire range between $\pm$\SI{1}{V} (see Supplementary Note~2 and Figure~S2). This result can lead to one of the three following implications: Firstly, the magnetization of the type III Fe ML reconstruction could be quenched. This can be excluded due to the residual spin contrast in the low magnetic field range and since this type of ML induces Shiba states in the band gap of the superconducting substrate (see below). Secondly, while the tip's magnetization is out-of-plane for $|B_z| >$~\SI{0.5}{T}, a strong in-plane magnetic anisotropy could force the Fe spins in the type III ML islands into the direction perpendicular to the tip magnetization, thus yielding zero spin contrast. This scenario can be excluded by Monte-Carlo simulations (see Supplementary Note~4 and Figure~S4) showing that, even for relatively strong in-plane magnetic anisotropies of a ferromagnetic island, the maximum magnetic field would still yield an out-of-plane magnetization of $25\%$ of the saturation magnetization which could be experimentally detected,~\cite{Doi2015} in conflict with the measurements. Therefore, the most likely scenario would be that the type III ML has a very low coercivity of <~\SI{0.4}{T} which is comparable to that of the tip.

In order to check the general validity of the above results for the different types of Fe ML islands, we investigated several islands of each type also using different tips (see Supplementary Note~5 and Table~1). They show, that the coercive fields do not vary considerably with island sizes or shapes, but are largely determined by the type of the reconstruction, i.e. dipolar interaction effects seem negligible. This can be rationalized by our Monte-Carlo simulations (see Supplementary Note~4 and Figure~S4)~\cite{Hagemeister2016, Hagemeister2020}. While this model simulates the very simplified situation of a ferromagnetic island with homogeneous properties, instead of a strongly reconstructed island where the magnetic properties will change from site to site, it still enables to understand the general trends by mapping the lateral changes in the magnetism in the different reconstructions to site-independent magnetic moments, onsite magnetic anisotropies and nearest neighbor Heisenberg exchange interactions. While we assume that the magnetic moment of the Fe atoms in the three types of MLs are similar ($\mu=2.5\mu_\textrm{B}$), we vary the magnetic anisotropy constants ($K_z, K_x=-0.05\ldots-0.5$~meV) and exchange constant ($J_1=1\ldots5$~meV per bond) within reasonable constraints. These simulations reveal that islands of the investigated sizes reverse their
magnetization by domain wall nucleation and propagation. The simulated coercive field is largely proportional to the square root of the product of anisotropy and exchange constants and can reproduce the measured coercive fields. Considering the differences in the coercive fields (\SI{2.3}{T} for type I, \SI{0.7}{T} for type II, and $<$~\SI{0.4}{T} for type III) we conclude that the overall anisotropy times exchange constants for the Fe atoms in the islands are decreasing by orders of magnitude with a ratio of $5.3 / 0.49 / <0.16$ going from type I, over type II, to type III ML. There might be, of course, additional effects of the different structural disorder of the three types of reconstructions on the coercive field via inducing nucleation centers for the domain walls. In the following, we will investigate the local properties of the Shiba bands which are expected to form by hybridization of the Yu-Shiba-Rusinov states of the Fe atoms in the islands~\cite{Roentynen2015,Roentynen2016,Li2016}, and how they differ for the three types of MLs due to their different structural and spin-dependent properties.

\subsection{Disordered Shiba bands in the reconstructed Fe MLs}
Spectroscopic line profiles taken with a superconducting tip from the Nb substrate across one of each type of the Fe ML islands are shown in Figure~\ref{fig3}a-c and the averaged spectra of those line profiles ontop of the Fe MLs and on Nb are shown in Figure~\ref{fig3}d. Since the tip has an energy gap of $\varDelta_\mathrm{t}$~=~\SI{1.00}{meV}, the energies of all observed features are shifted by \SI{1}{meV} relative to the Fermi energy $E_\textrm{F}$ (Methods). As we are interested in the states induced by the Fe islands in the gap of the substrate, we focus on the energy range from $\pm \varDelta_\mathrm{t} = \pm$~\SI{1}{meV} to $\pm (\varDelta_\mathrm{s} +\varDelta_\mathrm{t}) = \pm$~\SI{2.40}{meV}. While the substrate coherence peaks at $\pm (\varDelta_\mathrm{s} +\varDelta_\mathrm{t})$ are strongly suppressed ontop of all islands, there is an increase in spectral intensity all over the gap region compared to the Nb substrate. We assign this spectral intensity to the Shiba bands, which should form by hybridization of the Yu-Shiba-Rusinov states of the individual Fe atoms in the islands~\cite{Roentynen2015,Roentynen2016,Li2016}. Comparing the three types of MLs, there are pronounced differences in the overall intensities of the Shiba bands and their lateral variations which we will investigate in the following. While the coherence peak is increasingly suppressed going from the low coercivity type III over type II to the high coercivity type I islands, the energy-averaged intensity of the Shiba bands behaves opposite, i.e. it increases. This is also apparent from two-dimensional maps over an area with several islands (see Figure~\ref{fig4}a) of the spectral weight at $E_\textrm{F}$ (Figure~\ref{fig4}b) and at the coherence peak (Figure~\ref{fig4}c), respectively. Importantly, for all three ML types, the Shiba bands have significant intensities close to $E_\textrm{F}$, suggesting that their band widths are large enough to overlap with $E_\textrm{F}$. But there is no indication of a gap opening at $E_\textrm{F}$ (Figure~\ref{fig3}). However, this behaviour is complicated by a strong lateral variation in the spectral weight of the Shiba bands revealing shifts on the order of the substrate gap energy, which also differs between the three island types. The lateral variation is weakest for the type III ML and most pronounced for type I, and seems to have a correlation length comparable to the periodicity of the respective reconstruction. We assign these lateral variations to disorder in the Shiba bands induced by the strong structural disorder of the different Fe ML reconstructions. For the type I islands, a very strong and long range variation reveals nm sized areas ontop of the islands with an intense spectral weight at $\pm \varDelta_\mathrm{t}$ (see vertical white line on the center of the island in the spectral line profile of Figure~\ref{fig3}a), i.e. close to $E_\textrm{F}$, which are also visible in Figure~\ref{fig4}b. These special features occur only in sufficiently wide flat areas between the stripes when they have a minimum spacing of \SI{3}{nm} (see white dot in the center of the island marked on the red line on the STM image of Figure~\ref{fig3}a). However, note that the particle-hole asymmetry in the intensities of these features indicates, that they are not related to a zero energy state, but probably to a trivial state which is bound close to $E_\textrm{F}$ and localized by the disorder potential in the Shiba bands. Besides these different strengths of the disorder potentials in the Shiba bands for the three Fe ML reconstructions, the other main result of the current work is, that for all three types of ferromagnetic Fe MLs, there are no clear indications for any edge states, in the entire energy region of the substrate gap. Such edge states would be observed in the left and right regions of the spectral line profiles in Figure~\ref{fig3}a-c between the vertical white lines marking the edges of the islands. In particular, there are no indications for any zero energy edge modes as can be seen in Figure~\ref{fig4}b. As a side remark, we mention here, that the islands display a negligible inverse proximity effect on the Nb(110) substrate, which can be seen by the immediate recovery of the coherence peak spectral intensity in less than a nm distance to all islands edges (Figure~\ref{fig4}c,d), unlike it was observed for other systems.~\cite{Mougel2022}

\subsection{Conclusions}
In conclusion, our study unravels a correlation between different types of reconstruction-induced structural disorder in ferromagnetic Fe ML islands on clean Nb(110) with their coercive magnetic fields as well as with the spectral intensity and disorder in the Shiba bands which they induce in the superconductor. We assign the effect of the difference in reconstruction on the coercive field to a change in the magnetic anisotropy and/or exchange constant between the three different ML types. The different magnetic couplings in cooperation with the different types of structural disorder in the three ferromagnetic spin-lattices naturally explain a different disorder potential of their Shiba bands. The latter is quite massive, i.e. on the order of the superconducting energy gap for the spin-lattice with the strongest anisotropy times exchange constants and longest correlation length of the reconstruction, i.e. type I. Notably, the energetic position of YSR states has proven to be quite sensitive to disorder in previous studies, e.g. crucially depends on the exact adsorption site~\cite{Schneider2019}, on the position relative to oxygen impurities~\cite{Odobesko2020,Friedrich2021} or on the position relative to the charge density wave in $\mathrm{NbSe}_2$~\cite{Liebhaber2020}. The increasing spectral intensity in the Shiba bands for increasing anisotropy times exchange constants in the spin lattice naively seems intuitive, as in the classical model of Yu-Shiba-Rusinov, an increase in the magnetic moment or exchange interaction to the conduction electrons of the superconductor usually induces a shift of the Yu-Shiba-Rusinov states of individual atoms from the coherence peak into the gap region~\cite{Schneider2019, Kuester2021}. However, as this shift continues, the spectral weight will move towards the coherence peaks on the other side of $E_F$, which would probably balance the former effect. In the spin-lattice, in addition to these effects, an increasing exchange coupling between Yu-Shiba-Rusinov states will widen the Shiba bands,~\cite{Schneider2022} and this exchange coupling most probably is also different for the three ML types as it might be linked to the one that determines the coercivity. Most importantly, our study reveals that there are neither indications for a gap at $E_\textrm{F}$, nor of any edge states for all three types of disordered ferromagnetic spin-lattices investigated here, neither at nor off the Fermi energy. Naively thinking, the system might fulfill all requirements for topological superconductivity, i.e. a ferromagnetic spin-lattice which realizes Shiba bands that are overlapping with $E_\textrm{F}$ and very likely have a considerable spin-orbit interaction~\cite{Beck2021}. However, as shown in previous publications on 1D systems, the formation of topological edge modes can be hindered by the interaction of multiple Shiba bands originating from the five 3d orbitals of the Fe lattice~\cite{Schneider2021a, Kuester2021c}. Additionally, in the reconstructed ML system investigated here, we have experimentally proven a substantial disorder in the Shiba bands which can have both, a potential and a magnetic origin. As outlined in the introduction, excluding very specific and rare values of the chemical potential or specific types of disorder~\cite{Mascot2019b, Kezilebieke2022, Poeyhoenen2018}, the strong disorder on the order of the pairing energy present in the investigated lattices would destroy any topological gap as well as topological edge modes~\cite{Mascot2019b, Lu2020, Weststroem2016}. We, therefore, conclude that besides a specific design of the multi-orbital Shiba bands in ferromagnetic layers in direct contact to elemental superconductors,~\cite{Schneider2022} a pseudomorphic growth with the least amount of disorder is probably best suited for the design of two-dimensional topological superconductivity and the related Majorana edge modes. This yet adds another constraint and further narrows down the number of experimental systems which are suitable for the realization of this intriguing quantum state.

\section{Methods}\label{methods}
All measurements were obtained in two low-temperature ultra-high vacuum STM facilities, one is a home-built system operating at 6.5 K\cite{Wittneven1997}, and the other is a commercially available system with home-built UHV chambers operating at 4.5 K\cite{Loptien2014}. The Nb(110) crystal was cleaned by several cycles of \ch{Ar+} sputtering and high-temperature annealing using an electron beam heater at a power of at least 510~W (see Supplementary Note~1), following the recipe described by Odobesko et al.\cite{Odobesko2019}. Immediately after the cleaning, 0.52 ML Fe was deposited \emph{in-situ} from an e-beam evaporator at a rate of about one ML per minute onto the Nb(110) substrate held at room temperature (RT). STM-images were taken at constant tunnel current $I$ with a bias voltage $V$ applied to the sample, for which the corresponding values are given in the figure captions. d$I$/d$V$ point spectra were obtained by stabilizing the tip at a given point above the surface at stabilization current $I_\mathrm{stab}$~=~\SI{500}{pA} and bias $V_\mathrm{stab}$~=~\SI{1}{V}, switching off the feedback loop, and recording the differential tunneling conductance as a function of sample bias $V$ using standard Lock-In technique where the modulation voltage $V_\mathrm{mod}$~=~\SI{10}{mV} of frequency $f$~=~\SI{1.197}{kHz} is added to the bias voltage. Spectroscopic line profiles and spectroscopic fields were taken by recording point spectra on one- and two-dimensional grids, respectively, over the surface. In contrast, d$I$/d$V$ maps were recorded in constant-current mode in parallel to usual STM-images.

\begin{figure}
	\includegraphics[width=1\textwidth]{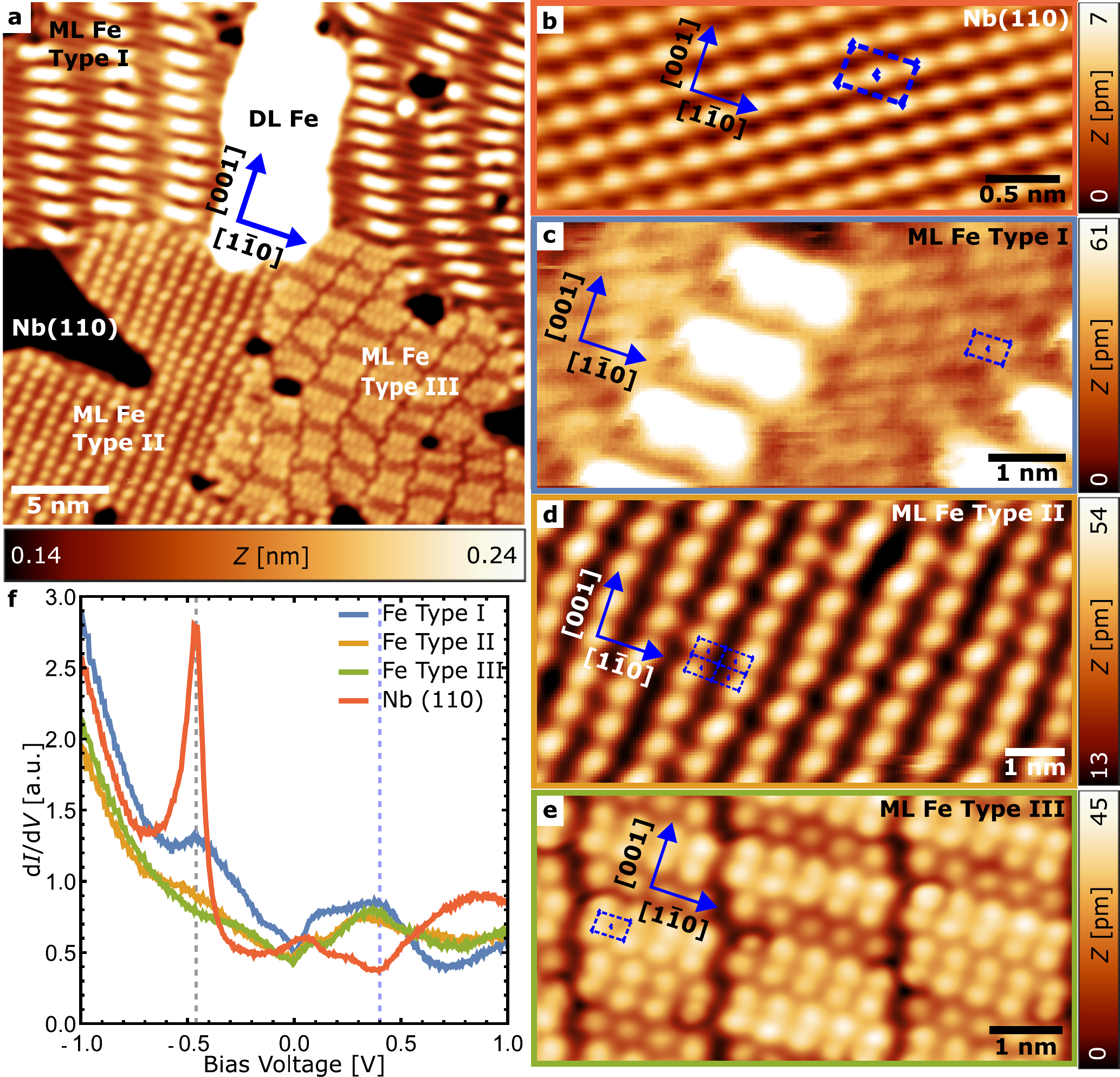}
	\caption{(\textbf{a}) STM image of ML Fe islands having the three distinct reconstructions I, II, III ($V$~=~\SI{-10}{mV}, $I$~=~\SI{4}{nA}, $T$~=~\SI{6.5}{K}). Furthermore, a DL Fe island can be identified as a bright stripe. (\textbf{b})-(\textbf{e}) Atomically resolved STM images of (\textbf{b}) the Nb(110) substrate and (\textbf{c}-\textbf{e}) of the individual Fe ML reconstructions ($T$~=~\SI{6.5}{K}; \textbf{b}: $V$~=~\SI{-10}{mV}, $I$~=~\SI{5}{nA}; \textbf{c}: $V$~=~\SI{-10}{mV}, $I$~=~\SI{5}{nA}; \textbf{d}: $V$~=~\SI{-10}{mV}, $I$~=~\SI{7}{nA}; \textbf{e}: $V$~=~\SI{-5}{mV}, $I$~=~\SI{100}{nA}). (\textbf{f}) Point spectra taken on the substrate and the respective Fe ML reconstructions as indicated. Vertical blue dashed lines mark the Nb(110) surface state at negative bias and a state of the Fe ML at positive bias ($V_\mathrm{stab}$~=~\SI{1}{V}, $V_\mathrm{mod}$~=~\SI{10}{mV}, $I_\mathrm{stab}$~=~\SI{0.5}{nA}, $T$~=~\SI{4.5}{K}).}
	\label{fig1}
\end{figure}

\begin{figure}
	\includegraphics[width=1\textwidth]{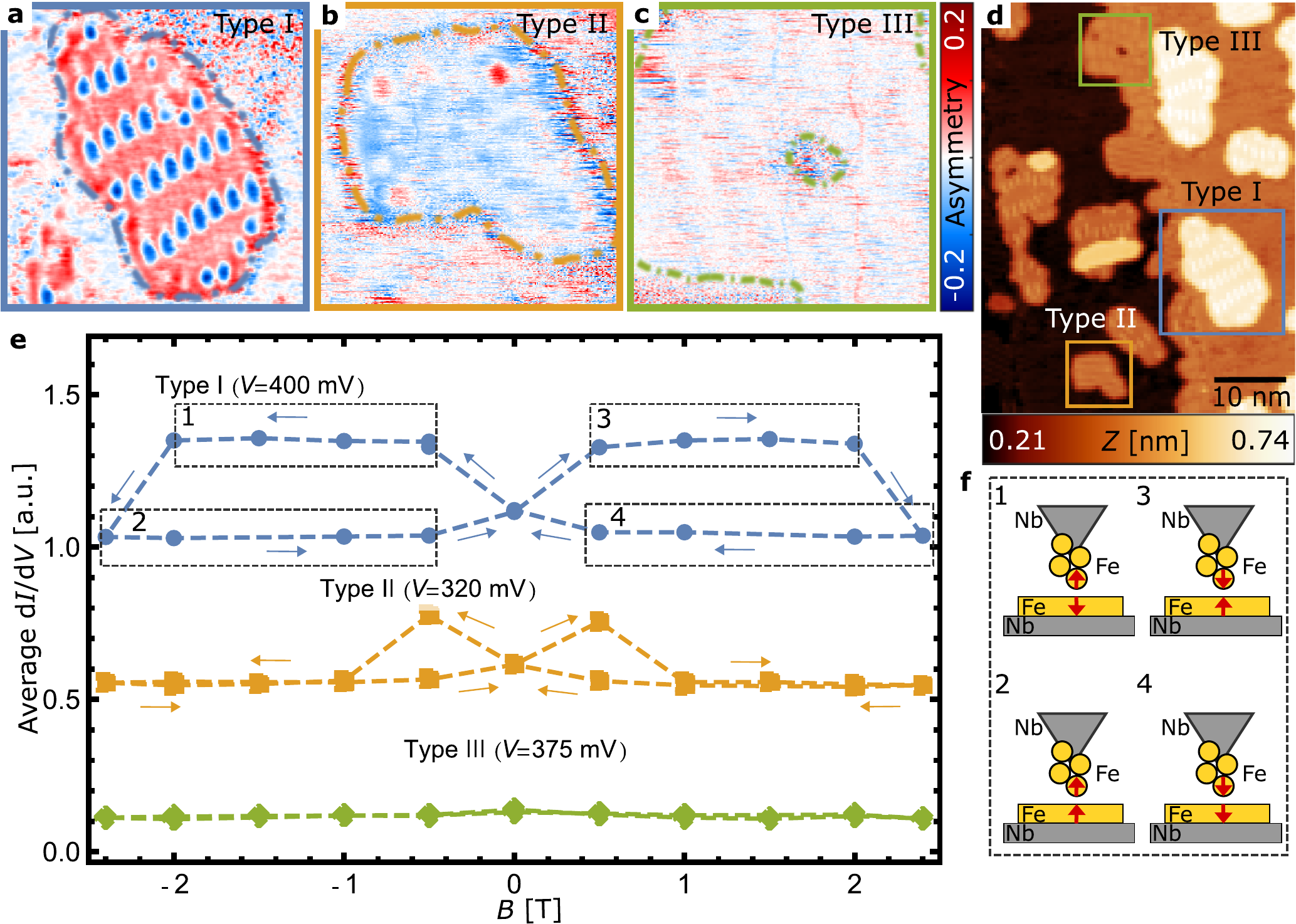}
	\caption{(\textbf{a})-(\textbf{c}) Spin asymmetry maps ($B_{z_1}$~=~\SI{-0.5}{T}, $B_{z_2}$~=~+\SI{0.5}{T}) of the different types of Fe ML reconstructions shown in the STM-image in (\textbf{d}) taken with the same spin-polarized STM tip ($I$~=~\SI{1}{nA}, $V_\mathrm{mod}$~=~\SI{50}{mV}, $V$~=~\SI{400}{mV} (a), $V$~=~\SI{320}{mV} (b), $V$~=~\SI{375}{mV} (c)). The outline of the individual islands is marked using dashed lines with a color according to the code. (\textbf{e}) Hysteresis loops of the same three islands, calculated from spin-resolved d$I$/d$V$ values at the indicated bias voltages averaged over selected areas of each of the different Fe ML reconstructions. (\textbf{f}) Exemplary sketch of the determined magnetizations of the tip and type I Fe ML island for different parts of the hysteresis loop as indicated by the according numbers in (e). Note, that the magnetic field dependence of the tip magnetization is the same for all three hysteresis loops.}
	\label{fig2}
\end{figure}

\begin{figure}
	\includegraphics[width=1\textwidth]{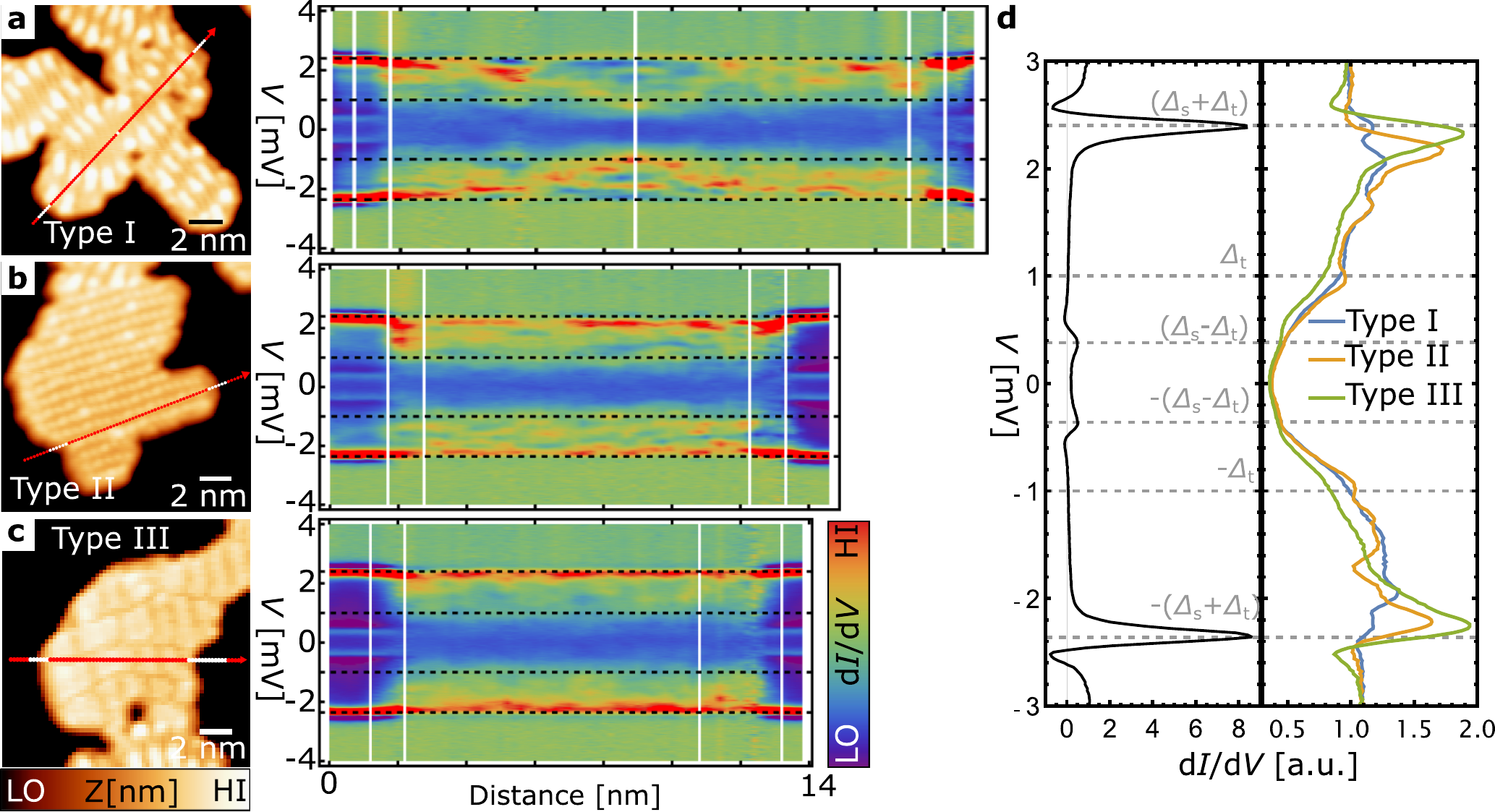}
	\caption{(\textbf{a})-(\textbf{c}) Left panels are STM-images of three Fe ML islands of each type of reconstruction as indicated ($I$~=~\SI{1}{nA}, $V$~=~\SI{-6}{mV}, $T$~=~\SI{4.5}{K}). Right panels are spectroscopic line profiles across each of the island types along the lines in the direction of the arrows ($I_\mathrm{stab}$~=~\SI{400}{pA}, $V_\mathrm{mod}$~=~\SI{0.1}{mV}, $V_\mathrm{stab}$~=~\SI{4}{mV}). The white dots on the arrows in the STM-images correspond to the positions where the spectra inbetween the white vertical lines of the spectroscopic line profiles have been taken. (\textbf{d}) Right panel: Spectra averaged ontop of the three islands from the spectroscopic line profiles in (a) to (c). Left panel: Spectrum averaged on an area of the bare Nb(110) surface. Gray or black dashed horizontal lines in the spectra are at $e \cdot V = \pm (\varDelta_\mathrm{t} - \varDelta_\mathrm{s})$, $e \cdot V= \pm \varDelta_\mathrm{t}$ and $e \cdot V = \pm (\varDelta_\mathrm{t} + \varDelta_\mathrm{s})$.}
	\label{fig3}
\end{figure}

\begin{figure}
	\includegraphics[width=1\textwidth]{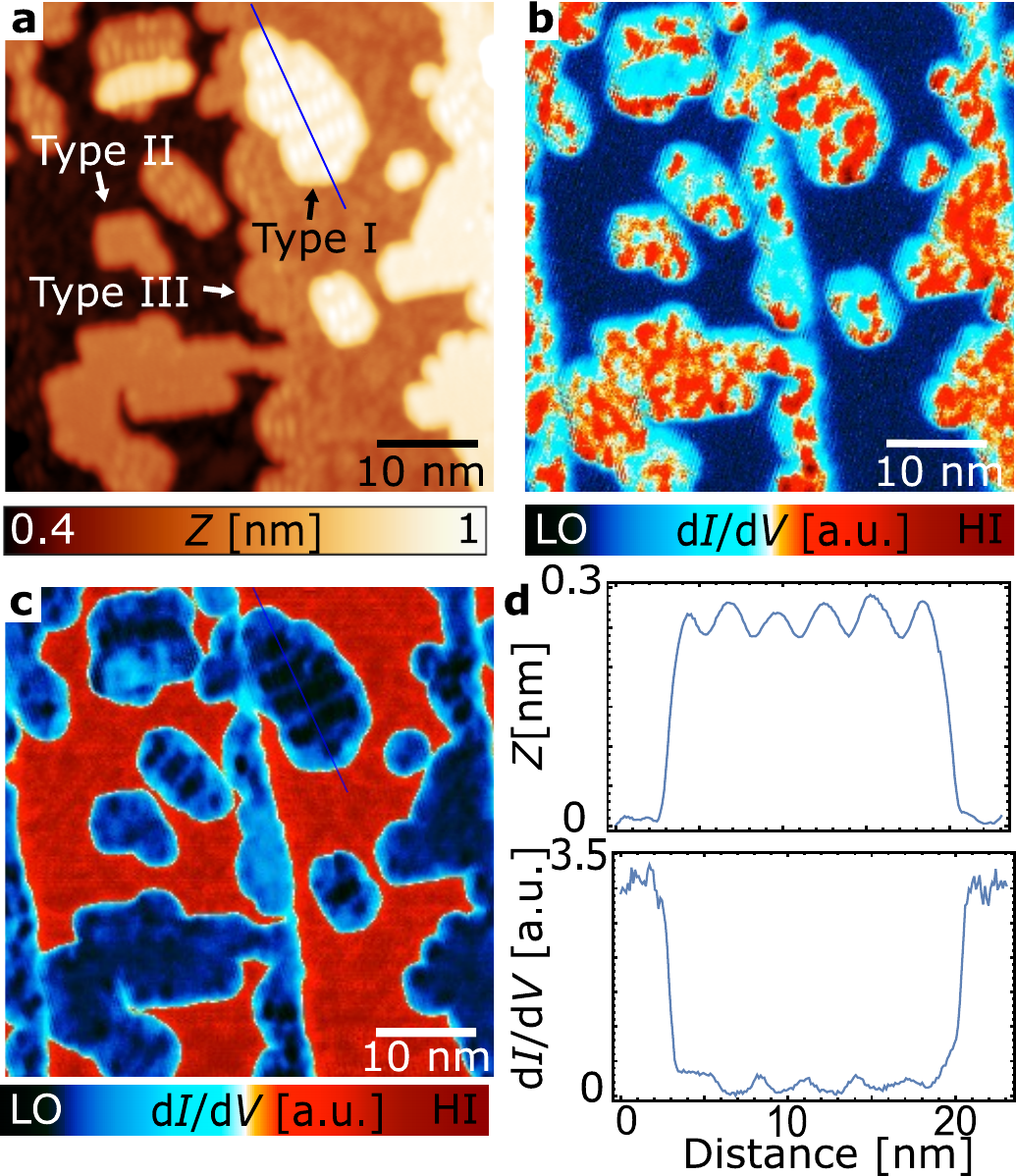}
\caption{(\textbf{a}) Overview STM image of an area with several Fe ML islands of all three types including some of the islands investigated in Figure~\ref{fig2}a-d. (\textbf{b}) Fermi energy and (\textbf{c}) Nb(110) coherence peak spectral weights, respectively, taken from spectroscopic grids over the same area recorded by following the tip height from (a) but with $e\cdot V$~=~\SI{1}{meV} = $\varDelta_\mathrm{t}$ (b) and $e\cdot V$~=~\SI{2.5}{meV} = $\varDelta_\mathrm{t}+\varDelta_\mathrm{s}$ (c). (\textbf{d}) Line profiles of the height (top panel) and Nb(110) coherence peak spectral weight (bottom) taken along identical lines across the type I island shown in (a) and (c), respectively. ($I$~=~\SI{200}{pA}, $V$~=~\SI{6}{mV}, $V_\mathrm{mod}$~=~\SI{0.1}{mV} (a); $V$~=~\SI{1.13}{mV}, $V_\mathrm{mod}$~=~\SI{0.1}{mV} (b); $V$~=~\SI{2.5}{mV}, $V_\mathrm{mod}$~=~\SI{0.1}{mV} (c)).}
	\label{fig4}
\end{figure}

For the atomically resolved STM-images (Figure~\ref{fig1}, $T$~=~\SI{6.5}{K}) an electrochemically etched tungsten tip was used. For the SP-STS measurements (Figure~\ref{fig2}, $T$~=~\SI{4.5}{K}), a mechanically sharpened Nb bulk tip was gently dipped into an Fe island in order to obtain a spin-polarized tip. After dipping, point spectroscopy on Nb(110) was performed to check for the presence of sub-gap states within the tip's superconducting gap, which indicate that magnetic material has been successfully transferred to the tip\cite{Huang2020,Schneider2021,Huang2021}. To emphasize the spin-resolved contrast, which is not easily seen in the individual d$I$/d$V$ maps $(\mathrm{d}I/\mathrm{d}V)_{ij}^{B_{z_1}}$ or $(\mathrm{d}I/\mathrm{d}V)_{ij}^{B_{z_2}}$ taken at various magnetic fields ${B_{z_1}}$ and ${B_{z_2}}$, respectively, spin asymmetry maps were calculated by \begin{equation}
	\label{Eq:1}
\textrm{asym}_{ij}= \frac{(\mathrm{d}I/\mathrm{d}V)_{ij}^{B_{z_1}}-(\mathrm{d}I/\mathrm{d}V)_{ij}^{B_{z_2}}}{(\mathrm{d}I/\mathrm{d}V)_{ij}^{B_{z_1}}+(\mathrm{d}I/\mathrm{d}V)_{ij}^{B_{z_2}}}.
\end{equation}
Furthermore, for the plot of the hysteresis loops in Fig 2 \textbf{e}, d$I$/d$V$ maps were taken for different out-of-plane applied magnetic fields varied in a loop sequence, as indicated. The d$I$/d$V$ values were normalized and averaged over a selected area of the different Fe island types and then plotted versus the external $B_z$ values.  All other spin-averaged measurements (Figures \ref{fig3} and \ref{fig4}, $T$~=~\SI{4.5}{K}) were performed with a pure Nb bulk tip, which increases the energy resolution at elevated temperatures beyond the Fermi-Dirac limit\cite{Pan1998,Ruby2015}. Therefore, the corresponding d$I$/d$V$ values reflecting features in the sample LDOS are shifted by the energy of the tip gap $\varDelta_\mathrm{t}$. The latter can be determined for each tip by taking a point spectrum on the substrate, see Figure~\ref{fig3}d. The peaks of largest intensity appear at $e\cdot V= \pm (\varDelta_\mathrm{s}-\varDelta_\mathrm{t})=\pm$~\SI{0.38}{meV} and $\pm (\varDelta_\mathrm{s}+\varDelta_\mathrm{t})=\pm$~\SI{2.38}{meV}. Thereby, we get the corresponding energy gaps of $\varDelta_\mathrm{t} =$~\SI{1.00}{meV} and $\varDelta_\mathrm{s} =$~\SI{1.38}{meV} (at $T$~=~\SI{4.5}{K}).

\begin{acknowledgement}
We thank Elena Y. Vedmedenko, Maciej Bazarnik, Eric Mascot as well as Roberto Lo Conte for fruitful discussions. J.J.G. and R.W. gratefully acknowledge funding by the European Union via the ERC Advanced Grant ADMIRE (No. 786020). L.S., D.W., J.W. and R.W. acknowledge funding by the Cluster of Excellence ‘Advanced Imaging of Matter’ (EXC 2056—project ID 390715994) of the Deutsche Forschungsgemeinschaft (DFG). K.T.T., Y.M., J.W. and R.W. acknowledge funding by the Deutsche Forschungsgemeinschaft (DFG, German Research Foundation) – SFB-925 – project 170620586.

\end{acknowledgement}


%\bibliography{bib}

\providecommand{\latin}[1]{#1}
\makeatletter
\providecommand{\doi}
  {\begingroup\let\do\@makeother\dospecials
  \catcode`\{=1 \catcode`\}=2 \doi@aux}
\providecommand{\doi@aux}[1]{\endgroup\texttt{#1}}
\makeatother
\providecommand*\mcitethebibliography{\thebibliography}
\csname @ifundefined\endcsname{endmcitethebibliography}
  {\let\endmcitethebibliography\endthebibliography}{}

\end{document}